\documentstyle[11pt,twoside,paspconf]{article}
\input psfig.sty

\begin{document}

%Invited paper presented at the workshop on ``Spectrophotometric Dating of Stars \& 
%Galaxies", held at Annapolis, MD, USA, in April 25-29, 1999

\title{Horizontal-Branch Stars as an Age Indicator}

\author{Young-Wook Lee, Suk-Jin Yoon, Hyun-chul Lee, \& Jong-Hak Woo}
\affil{Center for Space Astrophysics, Yonsei University, Seoul, South Korea}

\begin{abstract}
Surface temperature distribution of horizontal-branch (HB) stars is very sensitive to age 
in old stellar systems, which makes it an attractive age indicator. In this paper, we 
present the recent revision of our model calculations for the HB morphology of globular 
clusters. The result is more updated version of our earlier models (Lee, Demarque, \& 
Zinn 1994), which suggests that the HB morphology is more sensitive ($\sim$40\%) to age. We also 
present our new model calculations on the effect of HB stars in dating old stellar 
systems using the H$\beta$ index. Our results indicate that the effect of HB stars is 
rather strong, and suggest a possibility for the systematic difference in age between the 
globular clusters in the Milky Way Galaxy and those in giant elliptical galaxies. Finally, 
we compare these results with  the relative ages estimated from our far-UV dating 
technique for giant elliptical galaxies.
\end{abstract}

\section{Introduction}

For sufficiently old (t $>$ 8 Gyr) stellar populations, evolutionary models (e.g., Lee et al. 
1994, LDZ) predict that the surface T$_{e\!f\!f}$ distribution of horizontal-branch (HB) stars is 
very sensitive to age, which makes it an attractive age indicator. Because the HB stars 
are much brighter than main-sequence (MS) stars, the interpretation of HB morphology 
in terms of relative age differences would be of great value in the study of distant 
stellar populations where the MS turnoff is fainter than the detection limit. In the more 
distant stellar populations where even HB is fainter than the detection limit, HB stars 
still should have some crucial effect in spectrophotometric dating of old stellar 
populations based on integrated colors \& spectra. In this paper, we report our progress 
in the use of HB as an age indicator, both from the interpretation of HB morphology 
in the color-magnitude (CM) diagrams, and from the integrated spectra.

\section{Age as the Major Second Parameter}

Some five years ago, LDZ have concluded that age is the most natural candidate for 
the global second parameter, because other candidates can be ruled out from the 
observational evidence, while supporting evidence do exist for the age hypothesis. There 
is still much debate about this issue, and Table 1 summarizes the current situation. 
Although it may not be a complete list, it is clear 
from Table 1 \hfill that \hfill many \hfill pieces \hfill of 
\hfill supporting \hfill evidence \hfill still \hfill suggest \hfill that age is \hfill the

\begin{table}
\renewcommand{\baselinestretch}{1.05}
%\scriptsize
%\footnotesize
\caption{Age as the major 2nd parameter}
\small
\vspace{3mm}
\label{tbl-1}
%\begin{left}
\begin{tabular}{|l|l|l|}
\tableline
\multicolumn{3}{|c|}{\normalsize PROS}\\
\hline
Object & Author & Result \\
\tableline
Galactic globular    & Lee et al. (1988, 1994) 	&    Variation of age with HB \\
cluster  (GGC) & Sarajedini \& King (1989) 	&    type \ over a narrow range  \\
system  as a whole  & Chaboyer et al. (1992, 1996) & in [Fe/H]  \\
\tableline
NGC 288 & Bolte (1989) &  \\
vs. NGC 362 & Green \& Norris (1990) & $\Delta$t $\approx$ 2 $\sim$3 Gyr \\
    & Sarajedini \& Demarque (1990) &  \\
\tableline
Pal 12    & Gratton \& Ortolani (1988) & Considerably younger  \\
       & Stetson et al. (1989)    &   \\
\tableline
Rup 106 vs. & Buonanno et al. (1990) &   \\
M68/NGC6397 & Kubiak (1991)  & $\Delta$t $\approx$ 3$\sim$5 Gyr   \\
		& Da Costa et al. (1992)  &    \\
\tableline
	      & Buonanno et al. (1994) 	&   \\
Ter 7, Arp 2  &  Buonanno et al. (1995a,b) 	&  Considerably younger  \\
              &  Sarajedini \& Layden (1997)    &    \\
              &  Stetson et al. (1996)          &    \\
\tableline
IC 4499       & Ferraro et al. (1995)            & Considerably younger \\
              & Stetson et al. (1996)            &     \\
\tableline
Pal 1         & Rosenberg et al. (1998)          & Significantly younger  \\
              &                                  & than 47Tuc/M71   \\
\tableline
Pal 14 vs.  	& Sarajedini (1997) 		& $\Delta$t $\approx$ 3 $\sim$4 Gyr \\
NGC6752/M5	& 				&   \\
\tableline
Pal 3, Pal 4,   & Hesser et al. (1997)		& 1.5$\sim$3 Gyr younger \\
Eridanus	& Stetson et al. (1999)				& than M3/M5  \\
\tableline
M3 vs. M2	& Lee \& Carney (1999)		& $\Delta$t $\approx$ 2 Gyr  \\
\tableline
M3 vs. M13 	& Yim et al. (1999)		& $\Delta$t $\approx$ 2 Gyr  \\
\tableline
\end{tabular}

\vspace{5mm}

\begin{tabular}{|l|l|l|l|}
\tableline
\multicolumn{4}{|c|}{\normalsize CONS}\\
\tableline
Object	& Author & Argument & Our Comment \\
\tableline
GGC system 	& Stetson et al. (1996) 	&  			&   \\
as a whole 	& 				& 			&   \\
\cline{1-2}
	 	& VandenBerg et al. (1990)  	&   $\Delta$t is  	& New models  \\
		& Catelan \& Pacheco (1995)  	&   too small	  	& are more  \\
M3 vs. M13	& Ferraro et al. (1997) 	&	          	& sensitive  \\
		& Johnson \& Bolte (1998)	&   			& to age\\
		& Paltrinieri et al. (1998) 	& 			& \\
		& Davidge \& Courteau (1999) 	& 			& \\
\tableline
Draco Dwarf	& Grillmair et al. (1998)	& Older than M92 	& Very poor data \\
\tableline
Blue tail	& Buonanno et al. (1986)	& Internal 2nd 		& Most likely \\
\& Bimodal 	& Sosin et al. (1997)		& parameter 	& a local effect \\
HB  		& Rich et al. (1997)		& effect		& (see also text)   \\
 		& Ferraro et al. (1998)		& 			&  	\\
\tableline
\end{tabular}
%\end{left}
\end{table} 

\noindent
major global second parameter. Recent 
addition to this list include red HB clusters in the outer halo, such as Palomar 3 \& 4, 
and Eridanus from the {\it HST} photometry (Stetson et al. 1999), and Palomar 14 from the 
{\it WIYN} photometry (Sarajedini 1997). Before these observations, they used to say that 
these clusters will eventually solve the second parameter problem because they represent 
the most extreme cases of the second parameter effect. Now the observed age 
differences for these clusters, as estimated from the color difference method, are 
consistent with the age scenario of the 2nd parameter effect. Furthermore, J.-W. Lee \& 
Carney (1999) and Yim et al. (1999) recently report that their high quality observations 
of blue HB clusters M2 and M13 are  
consistent with the age hypothesis when compared to the redder HB clusters of 
similar metallicity, such as M3. 

        On the other hand, we see that there are now some papers in the literature
arguing that age may not be the major second parameter. Their argument, mostly based 
on M3-M13 pair, is that they did find some age difference, about 1$\sim$2 Gyr, but they 
think it is not enough to explain the difference in HB morphology. However, as 
described below, to within the observational errors, both in the age dating technique and 
in the HB type and metallicity measurements, our new models with the effects of recent 
developments are completely consistent with the observations. Other arguments 
include the {\it HST} photometry of Draco dwarf spheroidal galaxy (Grillmair et al. 1998), 
which has a red HB, in which they argue that Draco is not apparently younger than the 
blue HB clusters of similar metallicity. However, as they admit, the quality of their data 
is still very poor to reach any conclusion.
 
        We are not arguing that age can explain every aspect of the HB. In fact, LDZ 
also admitted that something else is also required to explain some peculiar features on 
the HB, such as long ``blue tail" and bimodal HB color distribution (see Table 1). 
However, these features are widely considerd to be a result of local effect, such as 
enhanced mass-loss in high density environment. Given the overwhelming supporting 
evidence for the age hypothesis, these local effects should be considered to be a third 
parameter effect, which is less important than the second parameter effect (see 
also Binney \& Merrifield 1998). Certainly, there is noise, and it is important to 
understand the origin of this third parameter effect in order to use the HB as a more 
reliable age indicator. In this respect, it is encouraging to see that recent observations 
with the Keck {\it HIRES} spectrograph (Behr et al. 1999) provide some compelling 
evidence that the ``blue tail" phenomenon is a natural result of abrupt diffusion 
mechanisms for stars hotter than 11,500K (see also Moehler et al. 1999). If, as 
suggested by these observations, the ``blue tail phenomenon" is a universal characteristic of 
clusters with stars hotter than 11,500K, then this phenomenon should not be 
considered to be a third parameter effect, since it is then rather a general feature of the 
very blue HB clusters. Furthermore, several lines of evidence now suggest that some 
globular clusters with peculiar CM diagram morphology, such as M54 (Sarajedini \& 
Layden 1995; Larson 1996) and $\omega$ Centauri (Lee et al. 1999), are nuclei of 
disrupted dwarf galaxies with internal age-metallicity relations. We suspect, therefore, a 
globular cluster with peculiar bimodal HB distribution, NGC 2808, may also represent 
such case. Note that it is among the most massive globular clusters in our Galaxy 
(Harris 1996). In any case, as demonstrated by Fusi Pecci et al. (1996), a procedure is 
available through which the measured HB morphology can be depurated from the effects 
of parameters (i.e., third parameter) other than age. So even in the clusters with 
peculiar HB morphology from unknown origin, a rough estimate of relative age is still 
possible from their HBs.

\section{Recent Developments \& New HB Population Models}

There are several recent developments that can potentially affect the relative age dating 
technique from HB morphology. The following effects are included in our most recent 
update of the LDZ HB models (Yoon \& Lee 1999): 

1. There is now a reason to believe that absolute age of the oldest Galactic globular 
clusters is reduced to about 12 Gyr, as suggested by new Hipparcos distance calibration 
and other improvements in stellar models (Gratton et al. 1997; Reid 1997; Chaboyer et 
al. 1998). As LDZ already demonstrated in their paper, this has a strong impact in the 
relative age estimation from HB morphology, because the variation of the red giant 
branch (RGB) tip mass (after the mass-loss, HB mass) is more sensitive to age at 
younger ages.

2. Now, we have new HB tracks with improved input physics (Yi et al. 1997) and 
corresponding new Yale isochrones (Demarque et al. 1996).
 
3. It is now well established that $\alpha$-elements are enhanced in halo populations. 
Specifically, we adopt [$\alpha$/Fe] = 0.4 for clusters with [Fe/H] $<$ -1.0, and thereafter 
we assume that it steadily declines to 0.0 at solar metallicity (e.g., Wheeler et al. 1989). In 
practice, the treatment suggested by Salaris et al. (1993) was used to simulate the effect 
of $\alpha$-elements enhancement.

4. Finally, Reimers'(1975) empirical mass-loss law suggests more mass-loss at larger 
ages. The result of this effect was also presented in LDZ, but unfortunately most 
widely used diagram (their Fig. 7) is the one based on fixed mass-loss.

        We found that all of the above effects make the HB morphology to be more 
sensitive to age (see Yoon \& Lee 1999 for more details). Therefore, as illustrated in 
Figure 1, now the required age difference is much reduced compared to Figure 7 of 
LDZ. Now, only 1.2 Gyr of age difference, rather than 2 Gyr, is enough to explain the 
systematic shift of the HB morphology between the inner and outer halo clusters. To 
within the observational uncertainties, age differences of about 1.5$\sim$2 Gyr are now enough 
to explain the observed differences in HB morphology between the remote halo clusters 
(Pal 3, Pal 4, Pal 14, \& Eridanus) and M3, and also between M3 and M13 (or M2). 
These values are consistent with the recent relative age datings both from the {\it HST} and 
high-quality ground-based data.

\section{Effect of HB on the Integrated Spectra}

If age is indeed the major parameter that controls HB morphology in addition to 
metallicity, then we expect some effect of it on the integrated spectra. For example, 
Figure 2 presents our models (Lee, Yoon, \& Lee 1999) that illustrate the effect of age 
sensitivity of the HB on the H$\beta$ index. First, we have compared in panel (a) our 
models without the HB with those of Worthey (1994) in order to make \hfill sure \hfill 
that \hfill our \hfill
calculations \hfill are \hfill consistent \hfill with \hfill previous \hfill investigations.

%\clearpage
%\newpage

\begin{figure}
\vspace{0.5cm}
\centerline{{\psfig{figure=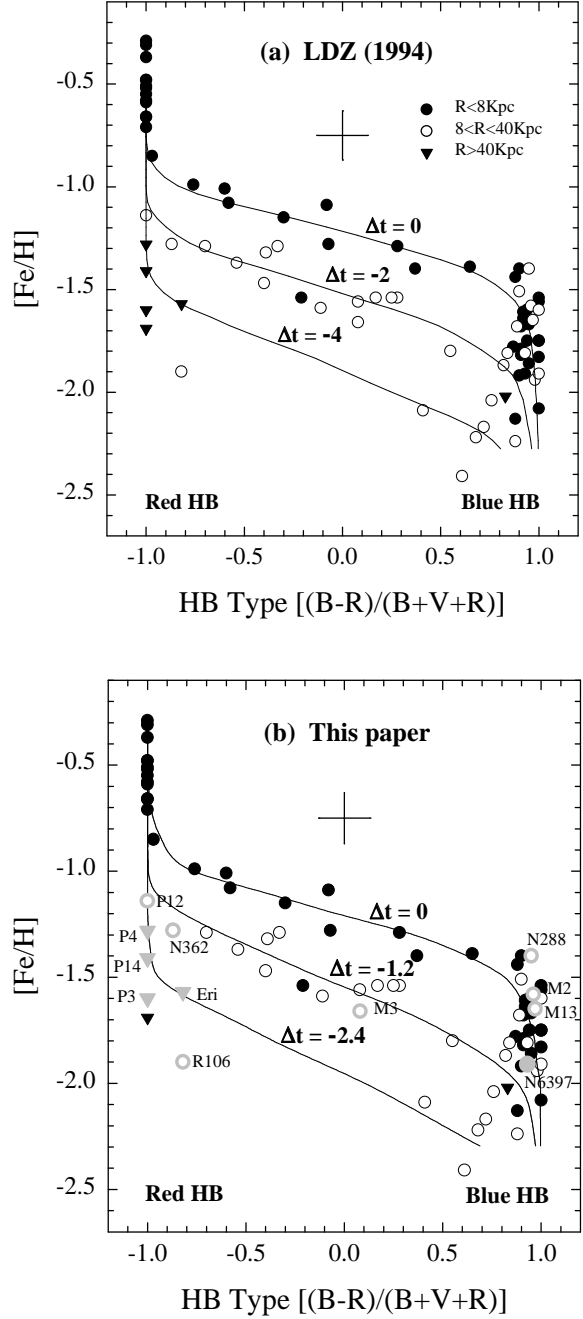,height=18cm}}}
\caption{Our new HB population models with the effects of recent developments (b) 
are more sensitive to age compared to our earlier models (a). $\Delta$t = 0 corresponds to the 
mean age of the inner halo (R $<$ 8 Kpc) clusters, and the relative ages are in Gyr.}
\label{fig-1}
\end{figure}

%\clearpage
%\newpage

\begin{figure}
\vspace{0.5cm}
\centerline{{\psfig{figure=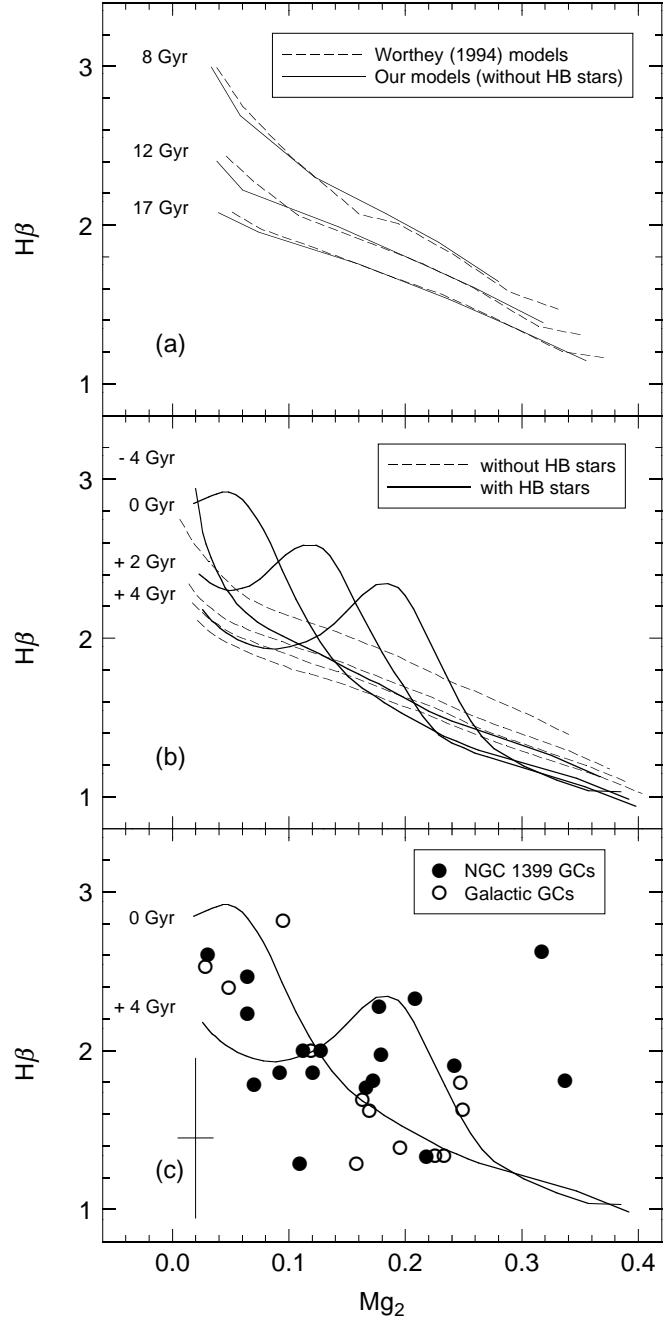,height=18cm}}}
\caption{Variation of H$\beta$ index with metallicity (Mg$_2$) and age for simple stellar 
populations (see text). In panel (c), observational data are compared with our models 
with HB, and the observational errors are indicated for clusters in NGC 1399.}
\label{fig-2}
\end{figure}

\clearpage
\newpage
\noindent
Thus, panel (a) illustrates only 
the effect of MS turnoff $T_{e\!f\!f}$ variation with metallicity and age. Confirmed that there is 
no systematic difference with previous investigations, we then added HB stars in our 
models, along the HB isochrones similar to those in Figure 1b, but including some very 
old ages ($\Delta$t = +2 \& +4 Gyr). The result is very striking as shown in panel (b). Of 
course, the wave-like features are due to the variation of the HB with metallicity and 
age. For given age, HB gets bluer with decreasing metallicity, and when the mean $T_{e\!f\!f}$ 
of the HB reaches around 9500K, the population models produce peak in the H$\beta$ 
index. But as metallicity decreases further, HB is becoming too hot to contribute 
significantly to the integrated H$\beta$ index.

        The advent of large ground-based telescopes making it possible to obtain some 
absorption features of globular clusters in nearby giant ellipticals so that our models can 
be compared directly with the observations. Data plotted in panel (c) of Figure 2 were 
obtained at Keck telescope by Kissler-Patig et al. (1998) and Cohen et al. (1998), using 
the same instrument, both for the globular clusters in the giant elliptical galaxy NGC 
1399 and those in the Milky Way Galaxy. At first glance, they may appear to be more 
or less the same, especially when considering still large errors in NGC 1399. However, 
some careful examination of the data indicates that there is some systematic difference in 
that the metal-rich clusters in NGC 1399 have higher H$\beta$ compared to the Galactic 
counterparts, while the opposite seems to be the case for more metal-poor clusters. 
Comparing this with our models, we can say that NGC 1399 system is probably several 
billion years older, in the mean, than the Galactic globular cluster system. Of course, 
given the large observational uncertainty, more observations are badly needed to confirm 
this possible scenario. If confirmed by future observations, this would indicate that the 
star formation in proto giant ellipticals started at an earlier epoch than in the less 
massive galaxies, such as our Milky Way Galaxy.   

        What is interesting to us is that a similar age difference is also inferred from 
the ``metal-poor HB solution" of the UV upturn phenomenon of giant ellipticals (Park \& 
Lee 1997; Yi et al. 1999). Our composite models (see also Lee et al., this volume) 
indicate that age difference of about 3 billion years can also explain the systematic 
difference in UV upturn between the giant ellipticals and the spiral bulges of the Local 
Group. Whether the UV upturn phenomenon is indeed a natural extension of the global 
second parameter effect observed in Galactic globular clusters (``metal-poor HB 
solution"), or it is rather due to the high $\Delta$Y/$\Delta$Z and enhanced mass-loss in super 
metal-rich population (``metal-rich HB solution") is still under debate. Fortunately, there 
are several observational tests for this problem that will eventually provide us more 
concrete calibration of the far-UV dating for old stellar populations. First, as pointed out 
by Lee (1994), the ``metal-rich HB solution" predicts many super metal-rich RR Lyrae 
stars in the Galactic bulge because super metal-rich HB stars must cross the instability 
strip as they move back to blue HB with increasing metallicity. In fact, detailed models 
by Lee \& Lee (1999) indicate that the expected number of super metal-rich RR Lyraes 
is compatible with more metal-poor ones actually observed in the bulge. Although 
extensive $\Delta S$ observations by Walker \& Terndrup (1991) found not a single super 
metal-rich RR Lyrae stars in this field, more efficient observations with the {\it Caby}
photometric system (Rey et al. 1999) would be useful to confirm this result. Second, 
{\it UIT} observations found strong radial gradients in UV colors within elliptical galaxies 
(O'Connell et al. 1992), and thus it is important to see whether they are correlated with 
internal metallicity gradients. Indeed, Ohl et al. (1998) report that they found no 
correlation between the far-UV gradients and internal metallicity gradients, from which 
they conclude that UV upturn phenomenon may not be simply related to overall metal 
abundances in galaxies. They even suggested age as an alternative, which is consistent 
with our ``metal-poor HB solution". Ongoing {\it STIS/HST} observations will provide more 
clear result on this problem. Finally, far-UV observations with the planned 
{\it GALEX} UV space facility will also provide useful database that is crucial in calibration 
of our UV dating technique for old stellar systems. Although the importance of new 
observations can not be overstated, it is already clear now that the detailed modeling of 
the HB is crucial in spectrophotometric dating of old stellar populations.

\end{document}